\documentclass[aps,prd,superscriptaddress,nofootinbib]{revtex4}   
\usepackage{amssymb}
\usepackage{amsmath}
\usepackage{amssymb}
\usepackage{graphicx}
\usepackage{subfigure}
\usepackage{color}
\usepackage{mathrsfs}
\usepackage{ulem}
\usepackage{booktabs}
\usepackage[dvipsnames]{xcolor}

\usepackage[breaklinks=true,colorlinks=true]{hyperref}
\hypersetup{colorlinks=true,citecolor=blue,linkcolor=blue,urlcolor=blue}

\usepackage[utf8]{inputenc}
\usepackage[english]{babel}

\begin{document}

\title{A resonance in phonons scattering off a kink in the absence of a Peierls-Nabarro potential}

\author{Danial Saadatmand\footnote{saadatmand.d@gmail.com}}
%\email{Saadatmand.d@gmail.com}
\affiliation{Alikhanyan National Laboratory (Yerevan Physics Institute), 2 Alikhanyan Brothers Street, Yerevan 0036, Armenia}
\affiliation{School of Mathematics and Statistics, Carleton University, Ottawa, Canada}
\author{Arpine Piloyan\footnote{arpine.piloyan@gmail.com}}
\affiliation{Alikhanyan National Laboratory (Yerevan Physics Institute), 2 Alikhanyan Brothers Street, Yerevan 0036, Armenia}
\author{David Amundsen\footnote{dave@math.carleton.ca}}
%\email{moradimarjaneh@iau.ac.ir}
\affiliation{School of Mathematics and Statistics, Carleton University, Ottawa, Canada}
\author{A. Moradi Marjaneh\footnote{moradimarjaneh@iau.ac.ir; moradimarjaneh@gmail.com}}
%\email{moradimarjaneh@iau.ac.ir}
\affiliation{Alikhanyan National Laboratory (Yerevan Physics Institute), 2 Alikhanyan Brothers Street, Yerevan 0036, Armenia}
\affiliation{Department of Physics, Qu.C., Islamic Azad University, Quchan, Iran}

\begin{abstract}

We investigate the interaction of small-amplitude waves called phonons, with an initially static kink in an exceptional discretization of the $\phi^4$ model that is free of the Peierls--Nabarro potential. Phonons are generated by a localized harmonic source and scattered from one side of the kink. By computing the transmission and reflection coefficients over the entire phonon band, we demonstrate that the scattering properties depend strongly on the lattice spacing. In the weak-discreteness regime ($h< 1$), the kink is nearly transparent and phonons are transmitted through it over most of the phonon spectrum. In contrast, for strong discreteness ($h>1$), significant reflection emerges even though the corresponding continuum $\phi^4$ kink is reflectionless. We further show that depending on the frequency of the incoming phonons, the kink experiences negative radiation pressure and is accelerated toward the incoming phonons for all lattice spacings considered, and this effect is much stronger for the strong discretness. The frequency dependence of the kink velocity and energy transfer is explained in terms of resonances associated with Doppler-shifted phonon frequencies and extrema of the phonon group velocity. Our results reveal that strong lattice discreteness can qualitatively modify phonon–kink interactions even in systems where the static Peierls–Nabarro potential is absent.

\end{abstract}


%11.10.Lm --- Nonlinear or nonlocal theories and models
%11.27.+d --- Extended classical solutions; cosmic strings, domain walls, texture,\\
%05.45.Yv --- Solitons,\\
%02.60.Cb --- Numerical simulation; solution of equations,\\
%02.30.Jr --- Partial differential equations,\\
%03.65.Pm --- Relativistic wave equations,\\
%03.50.-z --- Classical field theories,\\
%03.65.Ge --- Solutions of wave equations: bound states,\\

\maketitle

\section{Introduction}\label{sec:introduction}
The Klein-Gordon equation, in both the continuum and discrete forms, is widely used to describe properties of kinks (topological solitons) and their dynamical properties in a wide variety of physical systems, including crystal dislocations, ferroelectric domain walls \cite{Nataf:2020}, polyacetylene chains \cite{Ogata1:1986}, DNA denaturation dynamics \cite{Komarova:2005}, and mechanical metamaterials \cite{Zhou_2017, Zhang:2019,Lo_2021,PhysRevApplied.17.014004}. Owing to its rich nonlinear structure, the Klein-Gordon framework has also served as a paradigmatic model for investigating the interactions of kinks with other nonlinear excitations, impurities, and small-amplitude linear waves (phonons)  \cite{Braun2010-ux,Belova:1997bq, ASKARI2020109854, Azadeh.JHEP.2022, Javidan.PRE.2008, Hadipour.PLA.2020,Bai_2016,Abdelhady.IJMPA.2011, Saadatmand:2016pui}.

In continuum models such as the $\phi^4$ and sine-Gordon theories, phonons can transfer energy and momentum to kinks, excite their internal modes, and induce phenomena such as radiation pressure, resonant acceleration, and long-range transport \cite{Abdelhady.IJMPA.2011,Romanczukiewicz:2005jm,Forg_cs_2008}. In discrete systems, however, the situation becomes considerably richer due to the presence of lattice effects and the Peierls–Nabarro  potential (PNP), which can pin moving kinks, promote phonon radiation, and strongly modify the outcome of phonon–kink scattering processes \cite{PhysRevE.52.R2183}. 

The interaction between acoustic waves and mechanical kinks has been extensively investigated through theoretical and computational studies using canonical nonlinear models, most notably the discrete and continuum $\phi^4$ and sine-Gordon systems \cite{HASENFRATZ1977191, Theodorakopoulos, Ogata:1984,Abdelhady.IJMPA.2011}. These studies have provided significant insight into the mechanisms governing kink dynamics and their manipulation by external excitations. A major obstacle to the precise control of kinks, however, originates from the inherently discrete nature of the physical systems that support them, where the characteristic width of a kink is often comparable to the underlying lattice spacing. In such discrete settings, the breaking of the continuous translational invariance gives rise to the static  PNP barrier \cite{Peierls,Nabarro,PhysRevE.48.3077,Kevrekidis:2019saj}. The presence of this barrier severely limits kink mobility, causing propagating kinks to radiate energy in the form of phonons and eventually become trapped at lattice sites \cite{PEYRARD198488}. Consequently, special classes of discrete models, known as exceptional discretizations, have been developed in which the PNP barrier is completely absent \cite{Speight1997-hc,Dmitriev:2006qm}. 

More recently, the first experimental observation of
mechanical kink control and generation via acoustic wave packets including highly discrete kinks which eliminate the PNP barrier by supporting a zero-energy kink is reported  by the
use of an elastically-coupled KL chain \cite{Qian:2026}. Interaction of large-amplitude (anharmonic) phonons with the sine-Gordon kink has already been discussed \cite{Jaworski.PLA.1987}, where it was shown that in the limit of vanishing phonon amplitude the general solution reduces to the perturbative formula. The interaction of the phonons with the standing discrete breather in a one-dimensional chain with hard-type and soft-type anharmonicity has also been discussed \cite{Hadipour.PLA.2020}. It was demonstrated that for the case of hard-type anharmonicity (soft-type anharmonicity) solitons are more transparent for high frequencies (small frequencies) phonon waves, while they efficiently reflect small frequencies (high frequencies) phonons. A diode-like transport of the wave packets through a tilted discrete breather in the discrete nonlinear Schrödinger model with asymmetric on-site defect potential was investigated in \cite{Bai_2016}. This phenomenon can be observed in the form of optical beams in a waveguide array or it can also manifest as a Bose-Einstein condensate trapped in an optical lattice. The interaction of a $\phi^4$ kink with a wave packet in a trivial vacuum background has been studied in \cite{Abdelhady.IJMPA.2011}. For small and moderate amplitudes, the kink is displaced opposite to the propagation direction of the wave packet, and for larger amplitudes it gains kinetic energy, can be dragged by the wave packet, and particle production may occur.

Besides classical studies of phonon interactions, quantum scattering of radiation quanta, called mesons, has also attracted attention from physicists. In classical field theory radiation does not reflect off reflectionless kinks \cite{Romanczukiewicz:2005jm}, whereas in quantum field theory mesons can, in general, be reflected \cite{evslin2023elastickinkmesonscattering,Ogundipe:2025vdt}. In inelastic processes, the interaction may allow a meson to split into two mesons, and in the ultrarelativistic limit of the incoming meson a positive pressure is exerted on the kink \cite{evslin2022kinksmultiplymesons}. In the present work, however, we focus exclusively on the classical limit and demonstrate that, in the regime of strong discreteness, radiation (or phonons) can be reflected even by kinks that are reflectionless.

In the present study, we numerically analyze phonon scattering from a $\phi^4$ kink in a nonlinear Klein--Gordon chain free of the PNP. A lattice site located sufficiently far from the kink is driven harmonically to generate phonons in the chain. A small driving amplitude is employed to keep the generated excitations predominantly within the linear regime and to reduce possible nonlinear effects, such as resonant energy pumping, supratransmission, and pronounced phonon-induced kink acceleration
\cite{PhysRevE.96.042109,Evazzade2018-fi,Caputo_2001,Motcheyo2013}. Unlike our previous study of phonon scattering off an asymmetric kink \cite{Saadatmand:2023hqr}, where phonons could scatter from both sides, the present kink model is symmetric and we consider scattering only from one side of the kink.

 Two different discrete versions of the $\phi^4$ equations, free of the static PNP, were considered in \cite{Rakhmatullina2018-kz,Dmitriev:2006qm}\textbf{:} \textbf{o}ne of them is based on the conservation of energy, and the other one is based on the conservation of momentum. It has been demonstrated that discrete kinks can be highly mobile in this framework, such that they can be accelerated by any weak external force. The absence of the static PNP can be achieved by choosing a discretization which preserves the Bogomolny–Prasad–Sommerfield (BPS) on kink energy, so that a kink can move along the chain, practically radiating no energy if its velocity is very small, and its profile is not affected by the dynamical effects \cite{Speight1997-hc,Speight:1998uq,PhysRevE.76.026601}. A new prescription method, which is a realization of the BPS construction, has recently been introduced in \cite{Saadatmand:2023hqr}. The existence of sliding kinks, free of PNP, which travel at a constant velocity over a flat background without emitting any radiation in four discrete versions of the quartic-coupling theory has also been demonstrated \cite{Oxtoby:2005kh}.

This paper is organized as follows: In Sec.~\ref{sec:phi6model}, we describe the discrete version of the $\phi^4$ model and the main equations in the absence of the PNP; in Sec.~\ref{Phonon-kink}, a harmonic phonon source is introduced to generate phonons in the chain and numerical results for phonon–kink interaction are presented for the case where phonons originate from only one side of the kink. The effects of the Doppler shift and resonance on the kink velocity are discussed in Sec.~\ref{sec:resonance}, and our conclusions are presented in Sec.~\ref{sec:conclusions}.

\section{The model}\label{sec:phi6model}

We consider a real scalar field theory in $(1+1)$ dimensions. We first summarize the continuum formulation of the model and subsequently introduce its discretized realization. The continuum dynamics are described by the Lagrangian density
\begin{equation} \label{eq:lagrangy}
\mathcal{L}=\frac{1}{2}\dot{\phi}^2 -\frac{1}{2}\phi^{\prime2}-V(\phi),
\end{equation}
where $\dot{\phi} \equiv \partial_t \phi$ and $ \phi' \equiv \partial_x \phi $ denote temporal and spatial derivatives, respectively. We focus on the $\phi^4$ potential, which supports topological kink solutions, which has two degenerate vacua at $\phi\pm1$, allowing for topologically nontrivial field configurations that interpolate between these vacua. The simplest form of the potential is
\begin{equation}
V(\phi)=\frac{1}{2}(1-\phi^2)^2.
\label{eq:Phi4Potential}
\end{equation}
The dynamics is restricted to classical configurations governed by the Euler–Lagrange equation,
\begin{equation}
\ddot{\phi}-\phi^{\prime\prime}- 2\phi\left(1-\phi^2\right)=0.
\label{eq:Phi4EOM}
\end{equation}
For completeness, a detailed derivation of this equation from the continuum
Lagrangian density is provided in Appendix~\ref{app:derivation1}. This equation admits the well-known static kink and antikink solutions
\begin{equation}
\phi(x)=\pm \tanh(x).
\label{eq:Phi4Solution1}
\end{equation}
Here, the plus sign corresponds to the kink and the minus sign to the antikink. These topological solitons are related by the discrete symmetry $\phi \rightarrow -\phi$ and interpolate between degenerate vacuum states with identical local properties. Consequently, the kink exhibits symmetric exponential tails on both sides, with a decay rate determined by the mass of linear excitations around the vacuum. Because the asymptotic structure of the kink is identical as $x\to\pm\infty$, the scattering of phonons incident from the left or the right is expected to be equivalent. Therefore, throughout this work we consider phonons approaching the kink from only one side. This contrasts with the $\phi^6$ model studied in \cite{Saadatmand:2023hqr}, where the vacua are inequivalent and the kink possesses asymmetric tails, leading to direction-dependent scattering phenomena.

By linearizing the field equation around the static kink solution and considering small perturbations, one finds that the kink supports two localized bound states: the translational (or zero) mode with frequency $\omega = 0$, which arises from the translational invariance of the system, and the internal (shape) mode with frequency $\omega = \sqrt{3}$.

Of particular interest is the structure of the classical energy functional,
\begin{equation}
E_{\rm cl}[\phi]=\frac{1}{2}\int_{-\infty}^{+\infty}dx\,
\left[\phi^{\prime2}+(1-\phi^2)^2\right]
=\frac{1}{2}\int_{-\infty}^{+\infty}dx\,\left[\phi^\prime\pm(1-\phi^2)\right]^2
\mp\phi\left[1-\frac{1}{3}\phi^2\right]_{-\infty}^{\infty}.
\label{eq:BPS1}
\end{equation}
This functional admits a BPS decomposition, which allows one to derive first‑order equations whose solutions saturate a lower energy bound. This decomposition is crucial for constructing discrete versions of the model that are free of the PNP.

The two solutions in Eq.~(\ref{eq:Phi4Solution1}) are related by a sign flip and/or spatial reflection. Therefore, it is sufficient to consider a single BPS sector.
A BPS sector corresponds to a specific choice of sign in the first‑order equation that yields the minimal energy configuration for a given topological boundary condition. In practice, this first-order equation can be solved numerically by specifying a reference point $x_0$ and an initial condition $\phi(x_0)\in[-1,1]$. Fixing a BPS branch together with the field value at a single point uniquely determines the static solution.

To establish a framework suitable for a discretized formulation and to preserve the underlying BPS structure, we note that, up to surface contributions, the Hamilton functional can be written as
\begin{equation}
\mathcal{H}=\frac{1}{2}\int dx\left[\pi^2+u^2(\phi,\phi^\prime)\right],
\label{eq:Hamiltonian}
\end{equation}
where $\pi=\dot{\phi}$ and $u(\phi,\phi^\prime)=\phi^\prime-(1-\phi^2)$. We now introduce the discretized versions with an equi-distant lattice $x_n=nh$ where 
$n=-N,-N+1,\ldots,-1,0,1,\ldots,N$. Let us first consider the discretization introduced in \cite{Rakhmatullina2018-kz} which preserves the BPS property at the discrete level
\begin{equation}
u_n=\frac{\phi_n-\phi_{n-1}}{h}-
\left(1-\frac{\phi^2_{n-1}+\phi_{n-1}\phi_n+\phi^2_n}{3}\right).
\label{eq:discrete1}
\end{equation}
To perform numerical simulations, we introduce a discretized formulation of the model. By carefully choosing the discretization introduced in Eq.~(\ref{eq:discrete1}), we can eliminate the static PNP entirely, as shown by Speight \cite{Speight:1998uq, Speight1997-hc}.
In the continuum limit, the first term in $u_n$ reduces to $\phi^\prime$, while the second term approaches $-(1-\phi^2)$. The corresponding discretized field equations is then written as
\begin{eqnarray} \label{SpeightWardphi4}
\ddot{\phi}_n = \left(\frac{1}{h^2}+\frac{1}{3}\right)\left(\phi_{n-1}^{} - 2\phi_n^{} + \phi_{n+1}^{}\right) + 2\phi_n^{} - \frac{1}{9}\left[2\phi_n^3 + \left(\phi_n^{}+\phi_{n-1}^{}\right)^3 + \left(\phi_n^{}+\phi_{n+1}^{}\right)^3\right],
\end{eqnarray}
which has the following Hamiltonian
\begin{equation}
\mathcal{H}=\frac{h}{2} \sum\limits_{n=1-N}^N\left(\pi_n^2+u_n^2 \right)\,.
\label{eq:discretHamiltonian}
\end{equation}

Recall that   $u_n$  is defined as in Eq.~(\ref{eq:discrete1}). The condition  $u_n=0$  for every lattice site $n$ is the discrete analogue of the continuum BPS equation $\phi^\prime-(1-\phi^2)=0$. To generate a kink configuration, we determine a value of $\phi_n$ in the range $[-1,1]$ and compute $\phi_{n-1}$ from that equation. Similarly, we obtain $\phi_{n+1}$ from $u_{n+1}=0$ when $\phi_n$ is defined initially. Combining these two equations yields a quadratic algebraic relation between $\phi_{n-1}$, $\phi_n$, and $\phi_{n+1}$
\begin{equation}
\phi_{n\pm1}^2
+\left(\pm\frac{3}{h}+\phi_n\right)\phi_{n\pm1}
+\left(\phi_{n}^2\mp\frac{3}{h}\phi_{n}-3\right)=0\,.
\label{eq:Phi4kinksolution}
\end{equation}
Solving this quadratic equation for $\phi_{n\pm 1}$ gives
\begin{equation}
\phi_{n\pm1}^{} = -\frac{\phi_n^{}}{2} \mp \frac{3}{2h} \pm \frac{\sqrt{3}}{2}\sqrt{-\phi_n^2\pm\frac{6}{h}\phi_n^{}+\frac{3}{h^2}+4}.
\label{eq:Phi4kinksolution2}
\end{equation}
Choosing the appropriate values for $\phi_n$ ensures that each of these equations has one real solution and the field values remain in the interval $[-1,1]$ and approach the vacuum values as $n \to \pm \infty$. For $\phi_n=0$, we have an on-site kink and the inter-site kink is constructed for  $\phi_n=3/h-\sqrt{3+9/h^2}$. Iterating the procedure from an initial value $\lvert\phi_0\lvert<1$ reconstructs the entire discrete kink profile, which is free from the static PNP. We denote the resulting discretized kink solution by $\phi_n^{(0)}$.

\begin{figure*}[ht!]
\begin{center}
  \centering
  \subfigure[]
{\includegraphics[width=0.32
 \textwidth]{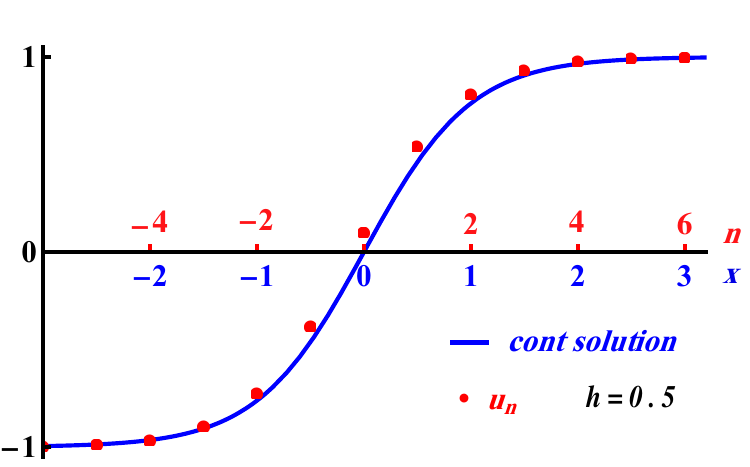}\label{fig:Phi4FieldAnalyticallyh05}}
\subfigure[]{\includegraphics[width=0.32
 \textwidth]{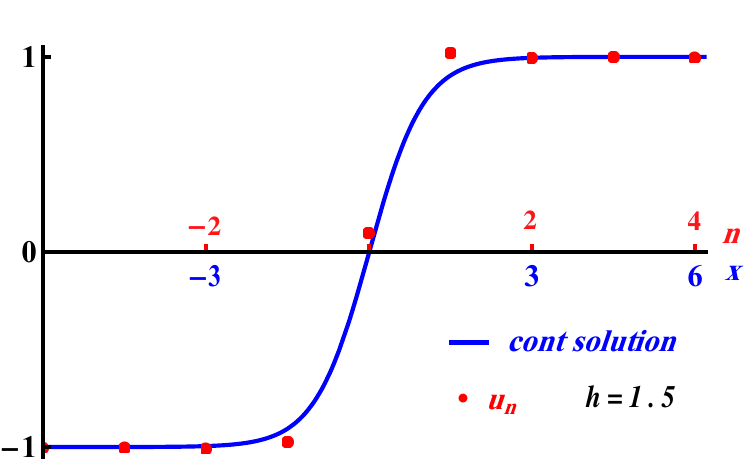}\label{fig:Phi4FieldAnalyticallyh15}}
\subfigure[]{\includegraphics[width=0.32
 \textwidth]{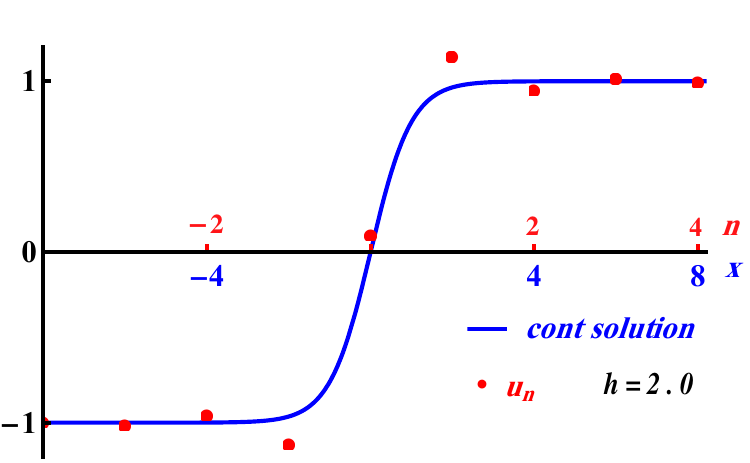}\label{fig:Phi4FieldAnalyticallyh20}}
\caption{Comparison of the continuum $\phi^4$ kink solution (solid blue line) and the discrete kink profile (red dots) for (a) (h=0.5), (b) (h=1.5), and (c) (h=2.0). As the lattice spacing increases, the discrete kink gradually deviates from the continuum solution and develops oscillatory tails around the vacuum values.}\label{fig:KinkProfile}
\end{center}
\end{figure*}

Examples of the static kink profiles constructed by iterating Eq.~(\ref{eq:Phi4kinksolution2}) are shown in Fig.~\ref{fig:KinkProfile} for several values of the lattice spacing,  $h=0.5, 1.5, 2.0$ with dots. The solid line represents the continuum kink solution given by Eq.~(\ref{eq:Phi4Solution1}). For  $h<1$, the kink approaches the vacuum monotonically, closely resembling the continuum solution. In contrast, for  $h>1$ the tails become oscillatory and approach the vacuum through damped spatial oscillations. This qualitative change in the asymptotic behavior is a characteristic feature of the discrete model at sufficiently large lattice spacings and reflects the increasing influence of lattice effects on the structure of the kink.

To study small-amplitude excitations, we linearize around the static kink background by introducing fluctuations $\epsilon_n(t)$ defined as deviations from the solution of Eq.~(\ref{SpeightWardphi4})
\begin{equation}
\phi_n(t)=\phi_n^{(0)}+\epsilon_n(t),
\label{eq:fluctuation}
\end{equation}
where $\epsilon_n \ll 1$, leading to the following linearized equation
\begin{eqnarray}
\ddot{\varepsilon}_n^{}&=&\frac{1}{h^2}\left(\varepsilon_{n-1}^{} -2\varepsilon_n^{} +\varepsilon_{n+1}^{}\right)+\frac{1}{3}\left[1-\left(\phi_n^0+\phi_{n-1}^0\right)^2\right]\varepsilon_{n-1}^{} \nonumber\\ &+&\frac{1}{3} \left[1-\left(\phi_n^0 +\phi_{n+1}^0\right)^2\right]\varepsilon_{n+1}^{}+\frac{1}{3}\left[4-2\left(\phi_n^0\right)^2-\left(\phi_n^0+\phi_{n-1}^0\right)^2-\left(\phi_n^0+\phi_{n+1}^0\right)^2\right]\varepsilon_n^{}.
\label{eq:Phi4Liniriazation}
\end{eqnarray}
In the linear regime, wave-like solutions are obtained using the standard ansatz for lattice vibrations, $\epsilon_n(t)\propto{\rm exp}\left[{\rm i}\left(nqh-\omega   t\right)\right]$. Substituting this expression into Eq.~(\ref{eq:Phi4Liniriazation}) yields the phonon spectrum around the static solution
\begin{equation}
\omega ^2(q,h)=4+4\left(\frac{ 1}{h^2}-1\right)\sin ^2\left(\frac{qh}{2}\right)
\label{eq:Omega}
\end{equation}
where $\omega$ denotes the frequency and $q$ the wave-number, both defined in dimensionless lattice units. The spectrum of the phonons lies between two lines $\omega^2_1=4.0$ and $\omega^2_2=4+4\frac{1-h^2}{h^2}$. For $h<1$, the width of the phonon band decreases as the lattice spacing $h$ increases. This can be seen explicitly from the dispersion relation for which the band edges are $\omega_{\min}=2$ and $\omega_{\max}=2/h$. 
The band width is $2(1/h-1)$, which decreases monotonically as $h$ approaches $1$ from below. 
At $h=1.0$, the phonon band collapses completely, and the dispersion relation becomes flat, with $\omega=2$ independent of the wave number $q$. Consequently, the group velocity vanishes everywhere in the Brillouin zone, $v_g=\partial\omega/\partial q=0$, indicating a fully dispersionless regime. In this limit, no propagating phonon packet can transport energy through the lattice, and the oscillations remain spatially localized. Similar flat-band behavior and localization effects in discrete Klein--Gordon lattices have been discussed in Ref.~\cite{Kakmeni2023Chaos}.

\begin{figure*}[ht!] 
\begin{center}
  \centering
 \subfigure[]
{\includegraphics[width=0.45
 \textwidth]{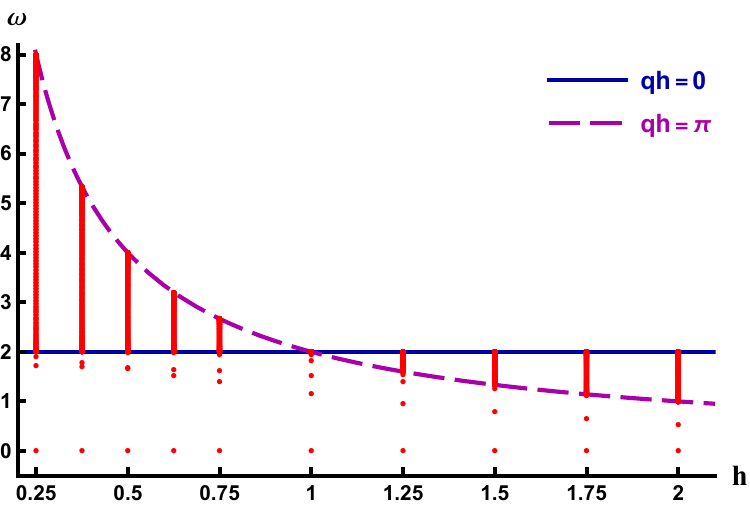}\label{fig:Omega_h}}
\subfigure[]
{\includegraphics[width=0.45
 \textwidth]{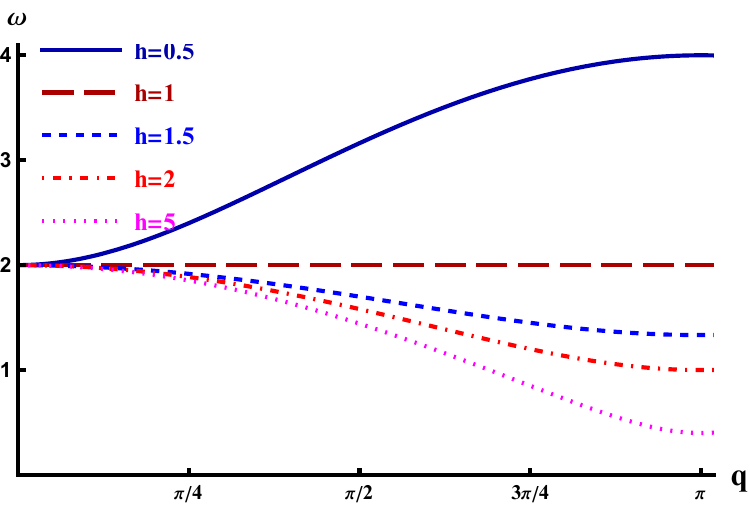}}
\caption{(a) Lower and upper edges of the phonon spectrum as functions of the lattice spacing $h$, obtained from Eq.~(\ref{eq:Omega}). (b) Dispersion relations of small-amplitude phonons for different values of  $h$, as given by Eq.~(\ref{eq:Omega}).}
\label{fig:Omega}
\end{center}
\end{figure*}

Conversely, for $h>1$, the band edges invert $\omega_{\min}=2/h$ and $\omega_{\max}=2$, giving a width $2(1-1/h)$, which increases with increasing $h$. 
This behavior is clearly seen in Fig.~\ref{fig:Omega}(a). The dispersion relation is also a function of the wave number $q$. For small lattice spacing, the frequency is an ascending function of wave number however for large lattice space is descending function of the wave number, as illustrated in Fig.~\ref{fig:Omega}(b).

As highlighted above, the dispersion relation depends explicitly on the wave number $q$. 
For $h<1$, the frequency is an ascending function of the wave number; however, for $h>1$, it is a descending function of the wave number. 
This behaviour follows directly from the group velocity,
\[
v_g(q,h) = \frac{d\omega}{dq} = \frac{1-h^2}{h\,\omega(q,h)} \sin(qh).
\]
Within the first Brillouin zone, the term $\sin(qh)$ is positive, so the sign of $v_g$ is determined solely by the factor $1-h^2$. 
Consequently, for $h<1$, the group velocity is positive and for $h>1$, we have negative group velocity.

To determine the eigenfrequencies of the discretized model, we solve the corresponding eigenvalue problem numerically for the static kink solution $\phi_n^{(0)}$. Most of the resulting eigenfrequencies lie within the phonon spectrum of the vacuum, while only a few are located below the lower edge of the phonon band. This is illustrated in Fig.~\ref{fig:Omega}(a), where the lower and upper boundaries of the phonon spectrum, given by Eq.~\eqref{eq:Omega}, are shown by the dashed and solid lines. The dots represent eigenfrequencies of modes that are spatially localized around the kink. These values are also collected in Table~\ref{tab:shape_modes_horizontal} for several values of $h$.

For all values of $h$, there exists a zero-frequency mode corresponding to the translational mode associated with the broken translational symmetry of the kink solution. This zero mode belongs to the full localized spectrum but is not listed in Table~\ref{tab:shape_modes_horizontal}, which contains only the nonzero internal, or shape, modes below the phonon band. The localized mode with the lowest nonzero frequency is the kink's internal mode. In the continuum limit, $h\to0$, its frequency approaches the well-known value $\omega=\sqrt{3}$, in agreement with the continuum $\phi^4$ theory. For completeness, the linear stability analysis of the continuum kink,
including the derivation of the translational zero mode and the internal shape mode with $\omega=\sqrt{3}$, is presented in Appendix~\ref{app:derivation2}. The number of entries in Table~\ref{tab:shape_modes_horizontal} therefore reflects the number of nonzero localized internal modes that remain spectrally separated from the phonon continuum for each value of $h$. For smaller lattice spacings, the system approaches the continuum $\phi^4$ limit, in which, apart from the translational zero mode of the full spectrum, only the standard internal shape mode lies below the continuum threshold. Additional discrete localized modes, when present, move toward the phonon-band edge as $h$ decreases and eventually merge with the continuum; hence they can no longer be identified as isolated bound states. For larger $h$, stronger discreteness modifies the effective spectral problem around the kink and can support additional localized modes below the phonon band, which explains the larger number of entries in the table.
\begin{table}[h]
\centering
\begin{tabular}{|c|c|c|c|c|c|c|c|c|}
\hline
$h$ & 0.1 & 0.25 & 0.5 & 0.75 & 1.25 & 1.5 & 1.75 & 2 \\
\hline
$\omega_1$ & 1.730 & 1.720 & 1.657 & 1.391 & 0.958 & 0.794 & 0.658 & 0.538 \\
\hline
$\omega_2$ & -- & -- & 1.681 & 1.614 & 1.392 & 1.257 & 1.126 & 0.999 \\
\hline
$\omega_3$ & -- & -- & -- & 1.954 & 1.549 & 1.316 & 1.140 & -- \\
\hline
\end{tabular}

\caption{The frequencies of the nonzero shape modes located below the phonon band for different values of the lattice spacing $h$. The translational zero mode is not included in the table. A dash indicates that no additional localized internal mode is spectrally separated from the phonon continuum for that value of $h$.}
\label{tab:shape_modes_horizontal}
\end{table}

%%%%%%%%%%%%%%%%%%%%%%%%%%%%%%%
\section{Phonon-kink Scattering}\label{Phonon-kink}

In this section, we investigate the interaction between propagating phonons and a static kink background for different driving frequencies and lattice spacings. We analyze two distinct regimes of the lattice spacing, namely $h<1$ and $h>1$, where the phonon dispersion undergoes a qualitative change. At the critical value $h=1$, the phonon band collapses to a flat dispersion with zero group velocity, while for $h<1$ and $h>1$ the system exhibits normal and anomalous dispersion, respectively. These distinct regimes give rise to markedly different phonon propagation and scattering properties. As will be shown below, these modifications have a significant impact on the transfer of energy and momentum between the incoming phonons and the kink.

To understand the physical implications of these dispersion regimes, it is instructive to establish an analogy with an electrical lattice. The discrete Klein--Gordon system studied here can be mapped onto a nonlinear transmission line (NLTL) composed of series inductors and shunt varactor diodes, where the discrete step size $h$ acts as a tuning parameter for the inter-site coupling impedance. As shown in related electrical network studies~\cite{Kakmeni2017CNSNS}, adjusting such coupling parameters can induce transitions between different transmission bands. Since the allowed phonon band shifts with $h$, a single fixed driving frequency $\Omega$ cannot propagate through the lattice across all values of $h$, as it would fall into the transmission gap (evanescent zone). Instead, by tracking frequencies $\Omega$ that remain within the allowed band for each regime, our simulations reveal a dual-mode behavior. For $h < 1$, the group velocity $v_g = \frac{\partial \omega}{\partial q}$ is positive, indicating a right-handed (normal dispersion) medium where wave packets propagate forward. For $h > 1$, however, the group velocity becomes negative ($v_g < 0$), transforming the lattice into a left-handed (anomalous dispersion) medium where the wave packet envelope propagates backward relative to the phase velocity. This left-/right-handed transition directly explains the reversal of the scattering dynamics presented in Fig.~3, where the direction of phonon reflection and transmission coefficients is reversed as $h$ crosses the critical value of $1.0$.

The interpretation of the negative group velocity in the anomalous regime $h>1$ requires some care. The negative value of $v_g$ refers to the positive signed wave-number branch. Since the dispersion relation is an even function of the signed wave number, $\omega(q,h)=\omega(-q,h)$, the group velocity is an odd function, $v_g(-q,h)=-v_g(q,h)$. Therefore, the two signed components $+q$ and $-q$ carry energy in opposite directions. In the scattering simulations considered below, the phonon source is placed on the left-hand side of the kink. The incident phonon packet observed in the simulations is therefore the component whose energy flux is directed from the source toward the kink, i.e., to the right in the present geometry. Hence, there is no contradiction between the negative group velocity associated with the positive-$q$ branch and the rightward propagation of energy in the simulations. In the $h>1$ regime, the energy-carrying envelope of the incident wave moves toward the kink, while the carrier phase propagates in the opposite direction, which is the physical meaning of left-handed propagation in the present lattice.

To generate a phonon wave packet that subsequently interacts with the kink, a single lattice site, $n=n^{\ast}$, is driven externally. Throughout all simulations, the total number of lattice sites is fixed at $N=800$. With the kink initially centered at $n=0$, the source is typically placed to its left at $n^{\ast}=-100$. To generate phonons with a characteristic frequency $\Omega$, a single lattice particle is subjected to an external harmonic driving of the form
\begin{equation}
\epsilon_{n^{\ast}}(t)=A\sin(\Omega t),
\label{DrivingForce}
\end{equation}
where $A$ denotes the driving amplitude. In all simulations presented here, we set $A=0.05$.

We have verified that qualitatively similar results are obtained for other sufficiently small values of the driving amplitude. In this regime, the induced excitations remain predominantly within the linear regime described by Eq.~(\ref{eq:fluctuation}), and the qualitative scattering behavior is therefore insensitive to the precise value of $A$. Moreover, placing the source sufficiently far from the kink reduces direct nonlinear excitation of the kink region. These choices substantially suppress nonlinear transmission effects but do not, in general, constitute an absolute exclusion of supratransmission, since enhanced energy transmission may also arise through nonlinear wave interactions, including wave-collision mechanisms~\cite{Motcheyo2013}.

We also verified that the qualitative scattering results are not sensitive to moderate changes in the position of the phonon source, provided that the source remains sufficiently far from the kink. In the small-amplitude regime considered here, the generated excitation becomes nearly monochromatic after the initial transient has passed. Therefore, changing the source-kink distance mainly introduces a phase shift in the wave reaching the kink, while the driving frequency and the time-averaged incident energy flux remain essentially unchanged. Although dispersion can affect the transient components generated during the switch-on stage, these transients are excluded from the time window used to compute the scattering coefficients. Since the transmission and reflection coefficients are evaluated from time-averaged energy measures after the incident, reflected, and transmitted waves are spatially separated, the reported scattering trends are robust with respect to the source position.

The resulting wave packet is characterized by its amplitude $A$ and driving frequency $\Omega$, and its subsequent interaction with the kink will be analyzed only as a function of the driving frequency $\Omega$. Since the source is localized at a single site, it emits phonons symmetrically; equal amounts of energy initially propagate toward the kink, i.e., to the right, and toward the nearest boundary, i.e., to the left. Consequently, the energy flow measured at the left boundary provides a direct and convenient measure of the energy carried by the incoming wave packet incident upon the kink prior to any scattering process. We denote this incident energy as $E_i$, which will be used later to define the transmission and reflection coefficients in Eq.~(\ref{eq:coefficients1}). The energy flow can be computed from the local energy density $e_n(t)$, while excluding the kink region from the integration in order to isolate the phonon contribution.

\begin{figure*}[ht!] 
\begin{center}
  \centering
{\includegraphics[width=0.49\textwidth]{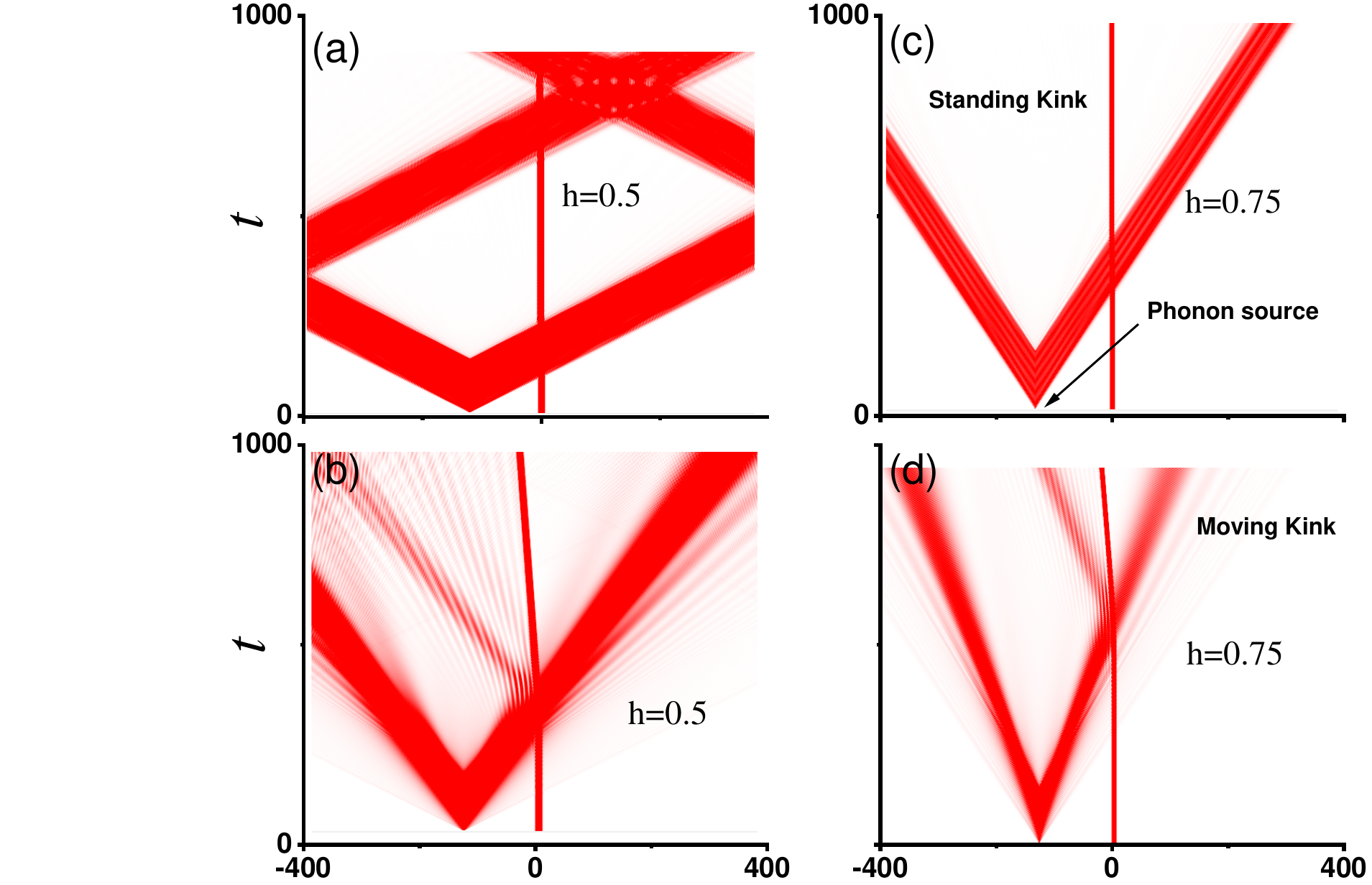}\label{fig:EnergyFlow1}}
{\includegraphics[width=0.49\textwidth]{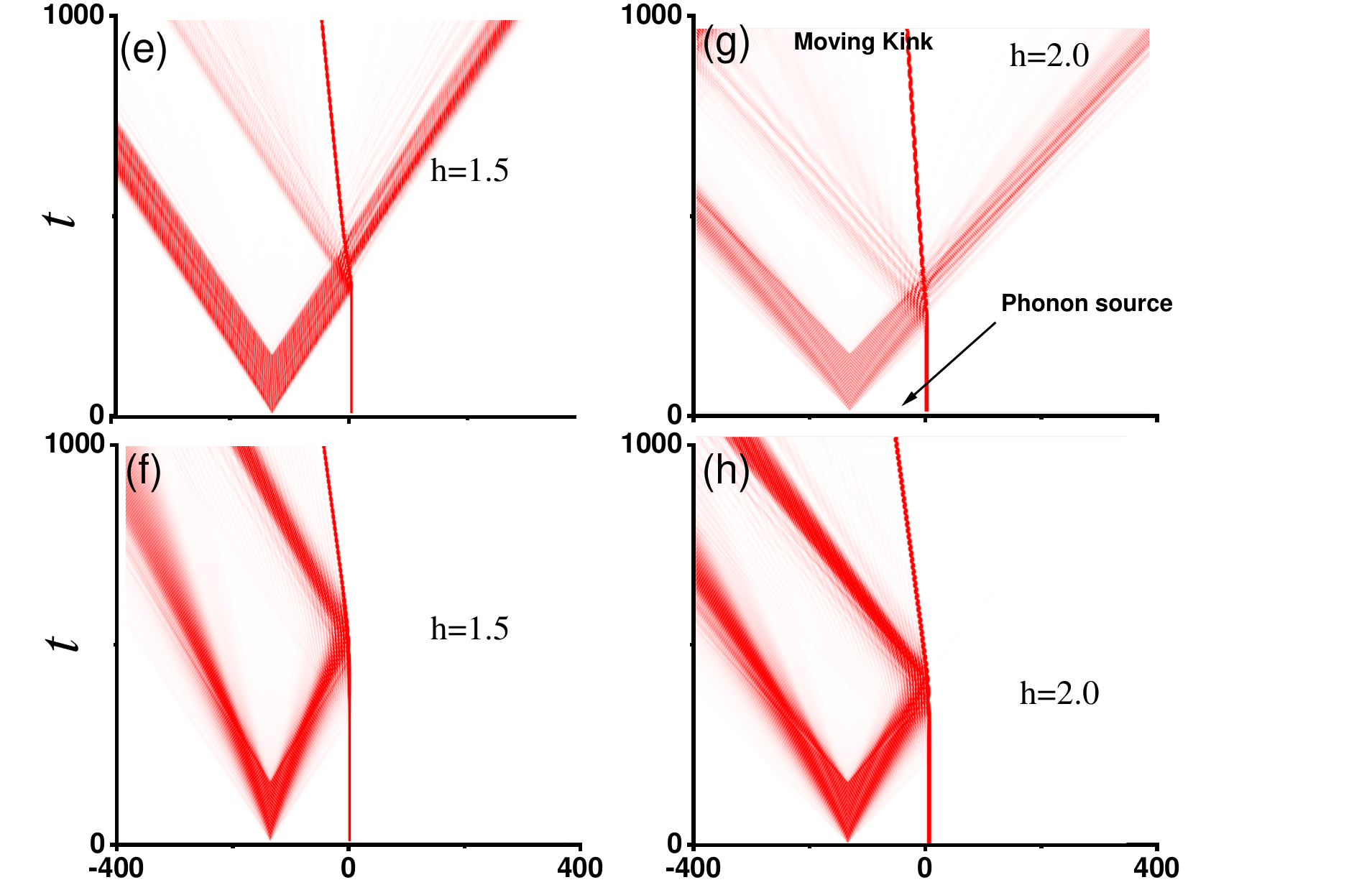}}
\caption{Total energy flow of all particles, together with the external energy source, for phonons incident from the left side of the kink. The driving frequencies are (a) $\Omega=3.1$, (b) $\Omega=2.3$, (c) $\Omega=3.9$, (d) $\Omega=2.6$, (e) $\Omega=1.6$, (f) $\Omega=1.2$, (g) $\Omega=1.9$, and (h) $\Omega=1.85$. The kink is initially located at the center of the lattice. The lattice spacing is $h=0.5$ in panels (a,b), $h=0.75$ in panels (c,d), $h=1.5$ in panels (e,f), and $h=2.0$ in panels (g,h).}
\label{fig:EnergyFlow}
\end{center}
\end{figure*}

To quantify the energy exchange between the incoming phonons and the kink, we first analyze the energy flow through the lattice for different driving frequencies and lattice spacings. This energy is calculated from the local energy density associated with each lattice site in the space-time evolution of the system as follows:
\begin{equation}\label{eq:EnergyPer}
e_n(t)=\frac{h}{2}\left(\dot{\phi}_n^2+u_n^2 \right),
\end{equation}
where all energy values below the threshold $e_n < 10^{-4}$ are discarded, and only those exceeding this cutoff are retained.

All panels in Fig.~\ref{fig:EnergyFlow} illustrate the time evolution of the total energy flow of all particles for several values of the lattice spacing $h$. It can be seen that in the regime of weak discreteness, $h < 1$, when phonons interact with the static kink, two different scenarios occur for the phonons: (i) full transmission and (ii) partial transmission accompanied by partial reflection. For the kink, two distinct outcomes are observed: it is either set into motion by the incoming phonons and acquires a negative velocity, corresponding to an effective negative radiation pressure, or it remains essentially at rest after the scattering. This can be seen in Figs.~\ref{fig:EnergyFlow}(a-d). However, in the regime of strong discreteness, $h > 1$, the kink is consistently accelerated toward the incoming phonons, indicating the presence of negative radiation pressure, as clearly shown in Figs.~\ref{fig:EnergyFlow}(e-h).

By comparing the energy propagating toward the kink with that propagating toward the boundary, one can directly assess the fraction of the incoming phonon energy that is transmitted and reflected by the kink. This analysis provides the basis for understanding the frequency-dependent scattering behavior of the kink with wave packets.

To make these results quantitative, we define the initial, reflected, and transmitted energies using the energy density in Eq.~(\ref{eq:EnergyPer}) as follows:
\begin{equation}\label{eq:coefficients1}
E_i=\frac{1}{2}\sum_{n=-N}^{-n_k} e_n(t^\prime)\,,\quad  
E_r=\sum_{n=-N}^{-n_k} e_n(t^{\prime\prime}) -E_i  
\quad {\rm and}\quad 
E_t=\sum_{n=n_k}^{N}  e_n(t^{\prime\prime}).
\end{equation}

Here, $t'$ is taken after the external driving has ceased but before the wave packet reaches the kink, while $t''$ is chosen after the scattering event, when the reflected and transmitted components are sufficiently separated in space. The kink contribution to the energy is excluded by introducing a spatial cutoff $n_k \approx 5$ around the kink center, ensuring that only phonons are included in the energy balance. The results are insensitive to moderate variations of these time windows and the cutoff, indicating that the decomposition is robust.

The quantities $E_i$, $E_r$, and $E_t$ are evaluated in the laboratory frame by integrating the energy density over spatial regions sufficiently far from the kink, after the incident, reflected, and transmitted phonon packets are clearly separated from the localized kink core. Thus, the coefficients defined from these energies represent the fractions of the incident phonon energy carried by the outgoing wave packets in the laboratory frame. Since the kink is accelerated during the scattering process, the interaction is not strictly elastic: part of the incident energy may be transferred to the translational kinetic energy of the kink or to its localized internal excitations. Consequently, an exact balance $E_i=E_r+E_t$ is not generally expected. The Doppler shift discussed below is used to interpret the resonance condition in the instantaneous moving-kink frame, whereas the energy coefficients are measured consistently in the laboratory frame.

\begin{figure*}[t!]
\begin{center}
  \centering
  \subfigure[]{\includegraphics[width=0.4\textwidth, height=0.2\textheight]{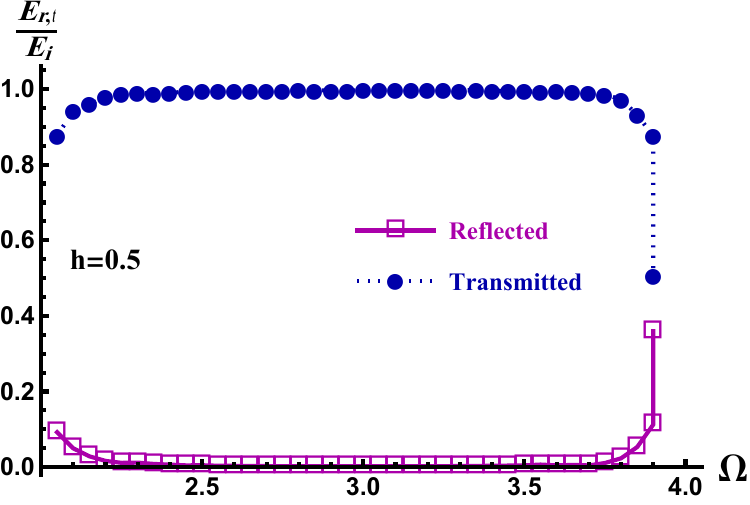}} 
  \subfigure[]{\includegraphics[width=0.4\textwidth, height=0.2\textheight]{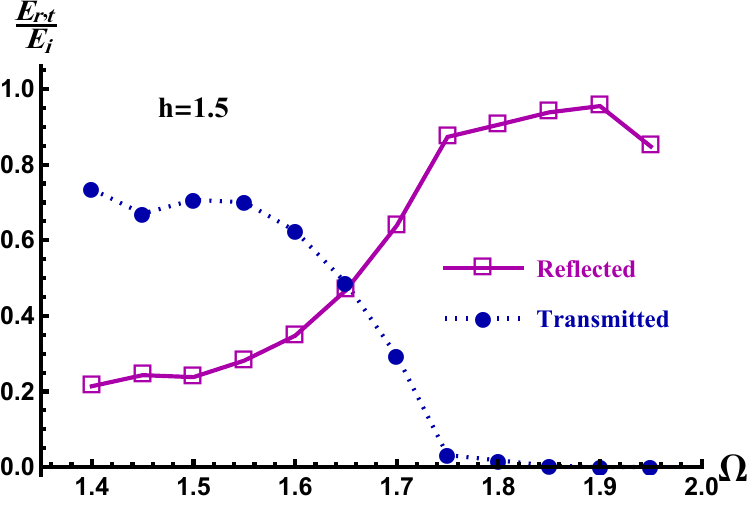}}
\\
  \subfigure[]{\includegraphics[width=0.4\textwidth, height=0.2\textheight]{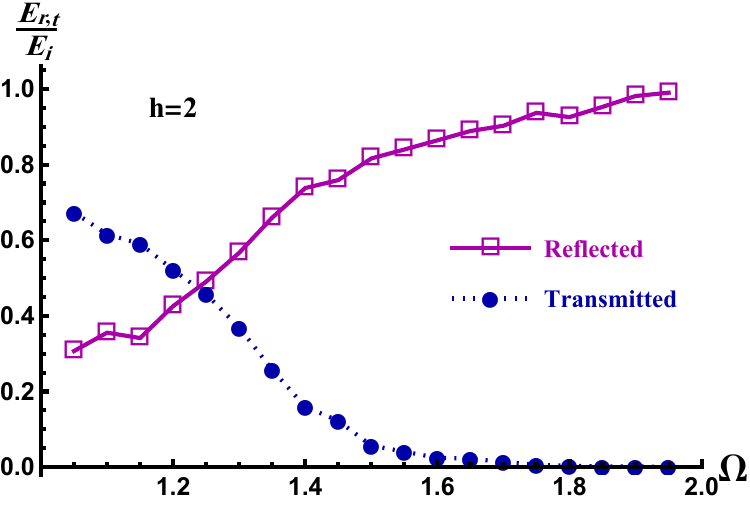}}
  \subfigure[]{\includegraphics[width=0.4\textwidth, height=0.2\textheight]{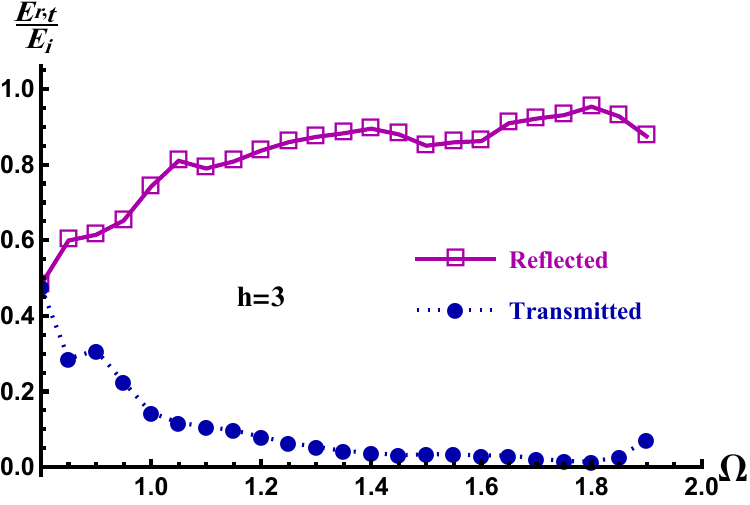}}
\end{center}
\caption{Transmission and reflection coefficients as functions of the driving frequency $\Omega$ for phonons scattering from the kink over the entire phonon band. The panels show the results for several values of the lattice spacing $h$, as indicated in each panel.}
\label{fig:EnergyTrans}
\end{figure*}

For the small amplitudes considered here, nonlinear effects remain negligible and the system is effectively in the linear scattering regime. In this regime, the residual energy transferred to the kink remains small compared with the total phonon energy, so the reflected energy is well approximated by $E_r \approx E_i-E_t$, consistent with Fig.~\ref{fig:EnergyTrans}.

It is also important to note that propagating phonon excitations exist only for driving frequencies $\Omega$ lying within the phonon bands shown in Fig.~\ref{fig:Omega}. For frequencies outside the phonon bands, the excitation remains spatially localized near the driven site and does not form a propagating mode capable of reaching the kink. This behavior is consistent with standard lattice vibration theory, where out-of-band driving leads to evanescent, exponentially decaying excitations rather than propagating phonon states. To ensure that the emitted wave packet indeed carries the same frequency as the external driving, we monitor the time evolution of one lattice site located far from the source. The time evolution of this site confirms that the frequency of the wave packet matches the driving frequency $\Omega$, indicating that the generated wave packet preserves the imposed temporal periodicity during propagation along the chain.

Figure~\ref{fig:EnergyTrans} shows the transmission and reflection coefficients as functions of the driving frequency $\Omega$ for different lattice spacings $h$. A clear qualitative distinction emerges between the regimes of weak and strong discreteness. For small lattice spacings, phonons are transmitted through the kink with high efficiency over most of the phonon band, indicating that the kink acts as a nearly transparent scattering center. Only within narrow frequency intervals close to the upper edge of the phonon band does the transmission decrease and partial reflection become noticeable.

In contrast, for larger lattice spacings, reflection becomes increasingly significant over a broad range of frequencies, even though the corresponding continuum kink is reflectionless. As the discreteness increases, the interaction between the propagating phonons and the localized kink modes is enhanced, leading to stronger scattering and a substantial reduction in transmission. Consequently, the kink behaves as a more effective barrier to phonon propagation, with a larger fraction of the incident energy being reflected back toward the source. For moderate discreteness, the phonons are partially transmitted and partially reflected by the kink, resulting in a mixed scattering regime in which the incident energy is distributed between the two outgoing channels.

\begin{figure*}[ht!] 
\begin{center}
  \centering
  \subfigure[]
{\includegraphics[width=0.45\textwidth]{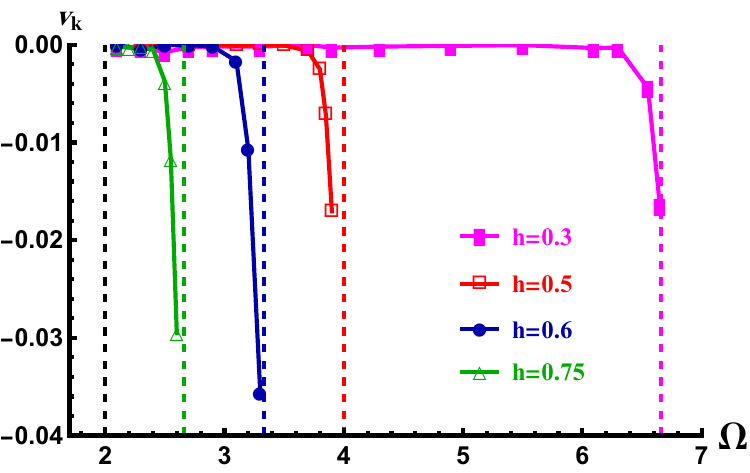}\label{fig:KinkVelocities2}}
  \subfigure[]{\includegraphics[width=0.45\textwidth]{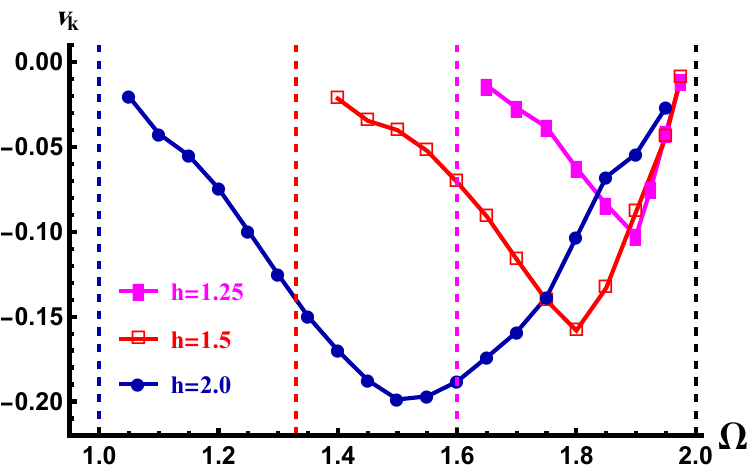}\label{fig:KinkVelocities1}}
  \caption{Kink velocity as a function of the driving frequency $\Omega$ for phonons scattering from the left side of the kink. (a) $h<1$ and (b) $h>1$. Note the different vertical scales. The vertical dotted line shows the boundary of the phonon band.}
\label{fig:KinkVelocity}
\end{center}
\end{figure*}

To calculate the velocity of the kink, we track its position in the chain at each time step and calculate the slope of its trajectory in the space-time graph. The results are shown in Fig.~\ref{fig:KinkVelocity} as a function of the driving frequency. In Fig.~\ref{fig:KinkVelocity}(a), the kink velocities are plotted in the region of weak discreteness. As can be seen, after the interaction of the phonons with the kink, the kink does not move and remains at rest for almost all frequencies of the phonon band, except for those close to the upper edge of the band. This behavior is plotted for several values of the lattice spacing. In Fig.~\ref{fig:KinkVelocity}(b), corresponding to the range of strong discreteness, the kink always moves toward the phonon source with velocities higher than those observed in the weak-discreteness case, as indicated by the different vertical scale. This occurs for all frequencies of the driving source inside the phonon band. The kink acquires its maximal velocity for frequencies in the middle of the band. We will explain the origin of this maximum in Sec.~\ref{sec:resonance} and demonstrate that it is caused by a resonance between the kink's internal mode and the Doppler-shifted frequency of the wave packet in the kink's rest frame.

\begin{figure*}[ht!] 
\begin{center}
  \centering
  \subfigure[]{\includegraphics[width=0.47\textwidth]{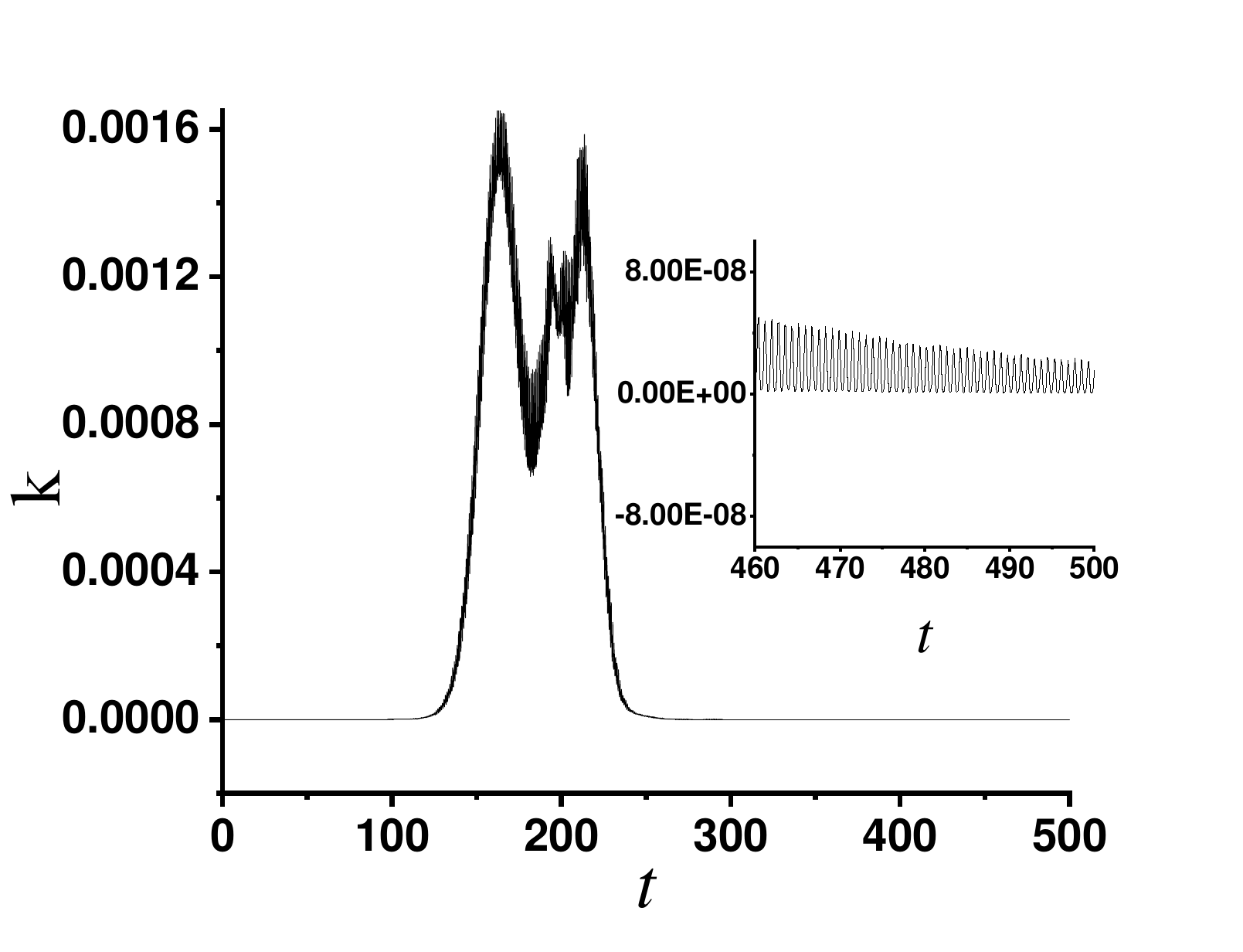}\label{fig:KineticEnergy1}}
  \subfigure[]{\includegraphics[width=0.47\textwidth]{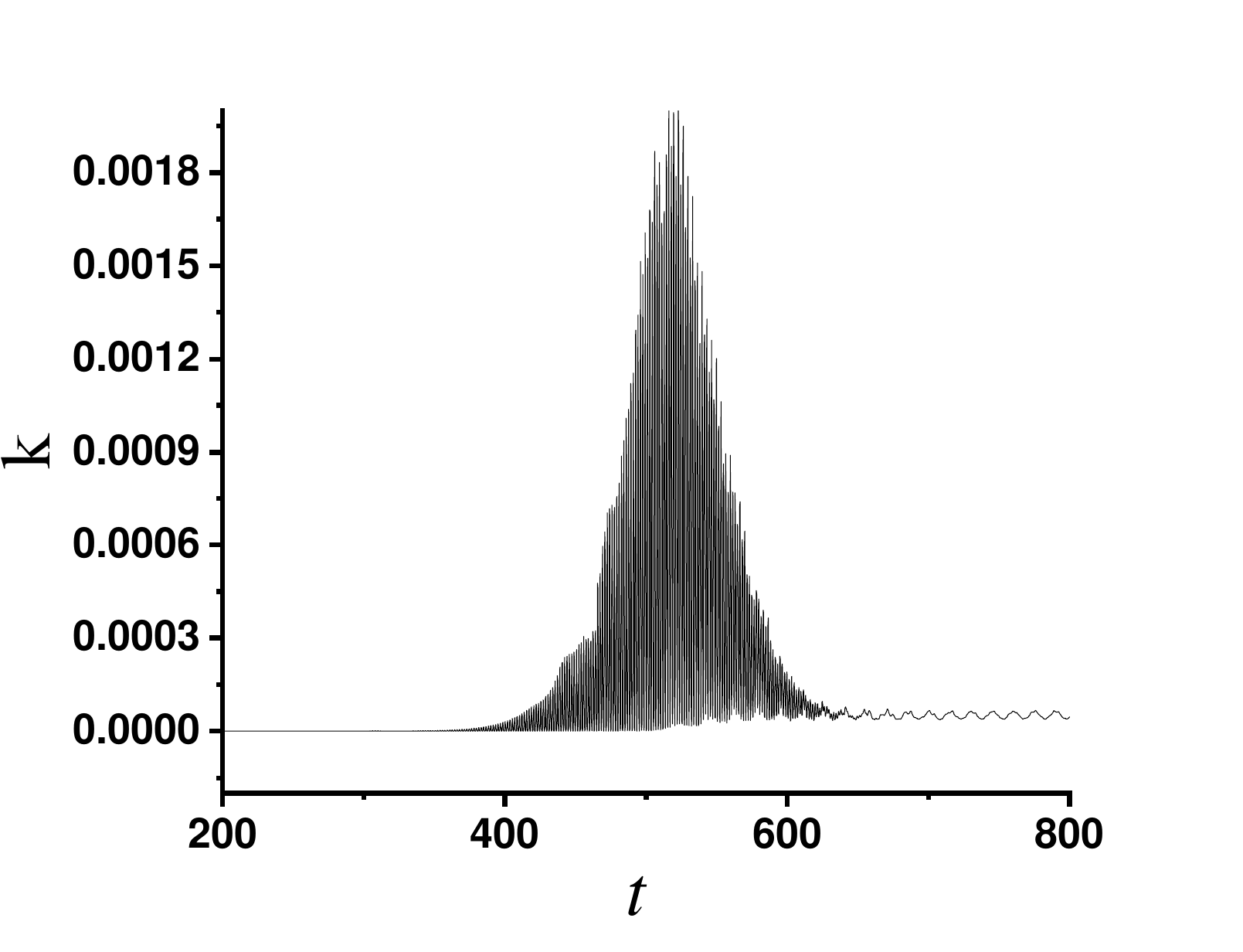}\label{fig:KineticEnergy2}}
  \\
  \subfigure[]{\includegraphics[width=0.47\textwidth]{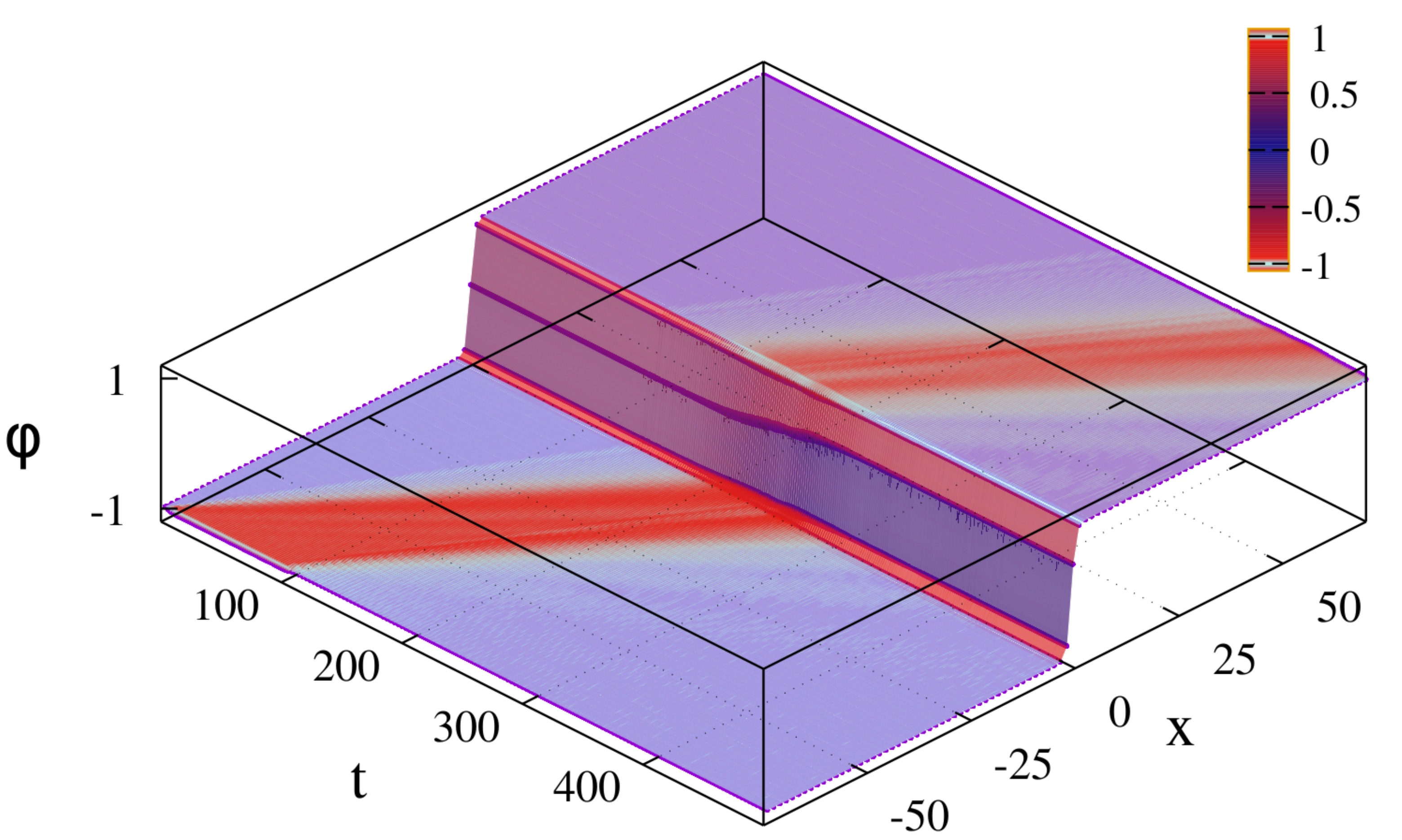}\label{fig:TXFh050Omega36}}
  \subfigure[]{\includegraphics[width=0.47\textwidth]{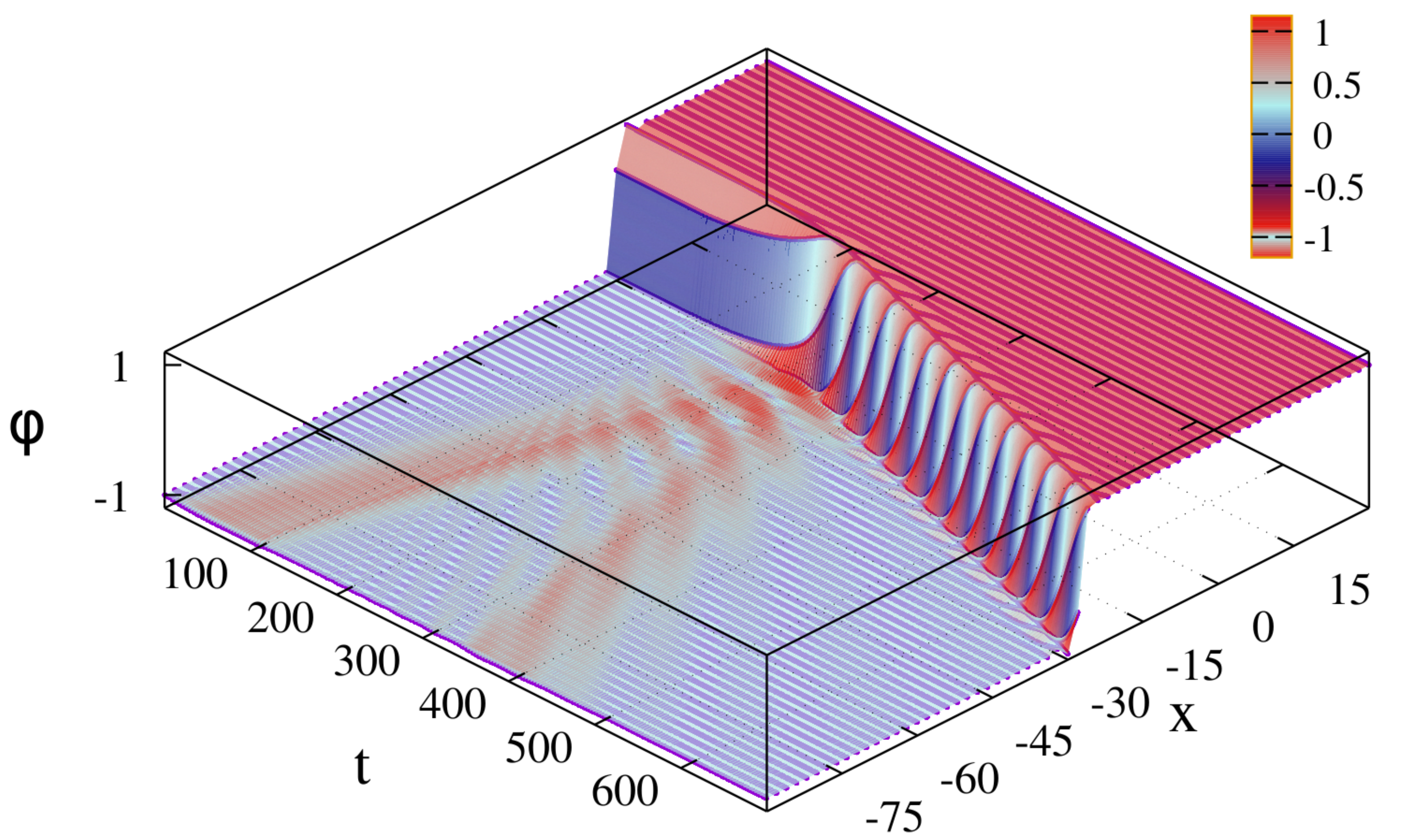}\label{fig:1}}
  \caption{Kinetic energy of the kink as a function of time. In panel (a), $\Omega=3.6$ and $h=0.5$, while in panel (b), $\Omega=1.9$ and $h=1.5$. The lower panels illustrate the 3D plots of the kink during the interaction with phonons.}
\label{KineticEnergy}
\end{center}
\end{figure*}

To better understand the kink dynamics in the weak- and strong-discreteness regimes, we plot the kinetic energy of the kink as a function of time. The results are presented in Fig.~\ref{KineticEnergy}. In panel (a), the lattice spacing is $h=0.5$ and the driving frequency is $\Omega=3.6$. It can be seen that the kinetic energy of the kink increases when the incoming phonon wave packet reaches the kink and subsequently returns close to its initial value after the phonons are transmitted through the kink. In addition, the amplitude of the kinetic-energy oscillations gradually decreases with time, see inset. This behavior indicates a continuous exchange of energy between the kink and the surrounding phonons. Such damping indicates that the frequency associated with the kink oscillations lies within the phonon band, allowing resonant coupling between the internal excitation of the kink and the propagating phonon modes. As a result, energy stored in the localized kink excitation is gradually radiated away into the lattice. In panel (b), we plot the kinetic energy of the kink for $h=1.5$ and driving frequency $\Omega=1.9$. In contrast to the weak-discreteness case, the kinetic energy exhibits persistent oscillations with an almost constant amplitude after the interaction between the phonon wave packet and the kink. The absence of noticeable damping indicates that the corresponding oscillation frequency of the kink lies outside the phonon band and therefore cannot efficiently couple to propagating phonon modes. As a result, the energy remains localized around the kink rather than being radiated into the lattice, leading to long-lived oscillations of the kink's internal excitation. The 3D plots of the kink motion during the interaction with phonons are also shown in the lower panels of each case to clearly illustrate this effect.

%%%%%%%%%%%%%%%%%%%%%%%%%%%%%%%%%%%%%%%%%%%%%%%%%%%%%
\section{Doppler Shifts and resonance on the Kink Velocity}\label{sec:resonance}

In this section, we investigate the effect of the Doppler shift on the phonon wave packet in the presence of a moving kink. We also seek to understand the origin of the extrema observed in the kink velocity and determine whether they are related to resonant interactions between the phonons and the kink's internal mode. 
% \textcolor{red}{Before discussing the Doppler correction, it is useful to clarify the sign convention used for the phonon wave number and the group velocity. The dispersion relation in Eq.~(\ref{eq:Omega}) is an even function of the signed wave number, $\omega(q,h)=\omega(-q,h)$, whereas the group velocity is an odd function, $v_g(-q,h)=-v_g(q,h)$. Therefore, the two signed components $+q$ and $-q$ carry energy in opposite directions. In the anomalous regime $h>1$, the positive signed wave-number branch, $0<q<\pi/h$, has a negative group velocity. This negative value indicates the reversal of the dispersion slope and is the signature of the left-handed character of the lattice. It should not be interpreted as meaning that the incident energy flux in the scattering simulations is directed away from the kink. In our simulations, the phonon source is a localized harmonic drive placed on the left-hand side of the kink. It fixes the driving frequency but does not impose a prescribed signed wave number. The incident phonon packet is the component whose energy flux is directed from the source toward the kink, i.e., to the right in the present geometry. Hence, in the $h>1$ regime, the energy-carrying envelope of the incident wave moves toward the kink, while the carrier phase propagates in the opposite direction. This is the physical meaning of left-handed propagation in the present lattice.}
We begin by considering the Doppler effect. In the rest frame of a kink moving with velocity $V_k$, a phonon with laboratory-frame frequency $\omega(q)$ and signed wave number $q$ is observed with a Doppler-shifted frequency. Using the signed wave number, this frequency can be written as
\begin{equation}\label{eq:DopPhBand}
\Omega(q)=\gamma\left(\omega(q)-qV_{k}\right),
\end{equation}
where
\[\gamma=\frac{1}{\sqrt{1-V_k^2}}\]
is the Lorentz factor. Equation~(\ref{eq:DopPhBand}) is written in terms of the signed wave number $q$. Therefore, the sign of the Doppler correction is determined by the relative signs of $q$ and $V_k$. Equivalently, if one uses the positive magnitude $Q=|q|$, the appropriate sign must be chosen according to the phonon branch under consideration. In the present left-handed regime, a wave packet carrying energy to the right may correspond to the signed branch $q<0$, because the group velocity and phase velocity have opposite signs.

To determine the frequencies at which the phonon group velocity $v_g$ has extrema, we eliminate the wave number $q$ from the dispersion relation, Eq.~(\ref{eq:Omega}),
\begin{equation}\label{eq:sinq}
\sin^{2}\frac{qh}{2}=\frac{\omega^2(q,h)-4}{4(\frac{1}{h^{2}}-1)},
\end{equation}
and express $v_g$ directly as a function of $\omega$:
\begin{equation}\label{eq:vgomega}
v_{g}
=
\pm \frac{1}{2 \omega }
\sqrt{\left(4-\omega ^2\right) \left(h^2 \omega ^2-4\right)}.
\end{equation}
The two signs in Eq.~(\ref{eq:vgomega}) correspond to the two signed wave-number branches. For $h>1$, the allowed phonon band is $2/h<\omega<2$. On the positive signed branch, $0<q<\pi/h$, the group velocity is negative, whereas the opposite branch has positive group velocity. Thus, when Eq.~(\ref{eq:vgomega}) is plotted with the negative sign for $h>1$, the plot represents the positive-$q$ branch and demonstrates the anomalous dispersion. It does not imply that the incident energy in the scattering simulations propagates to the left. The incident packet in the simulations is the branch whose group velocity, and hence energy flux, is directed from the source toward the kink.

\begin{figure*}[ht!] 
\begin{center}
  \centering
  \subfigure[]{\includegraphics[width=0.45\textwidth]{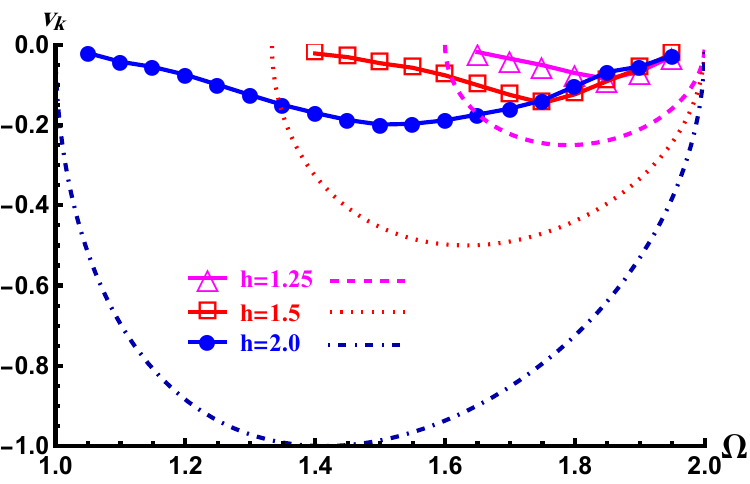}\label{fig:group_vel_Kinks_vel}}
    \subfigure[]{\includegraphics[width=0.45\textwidth]{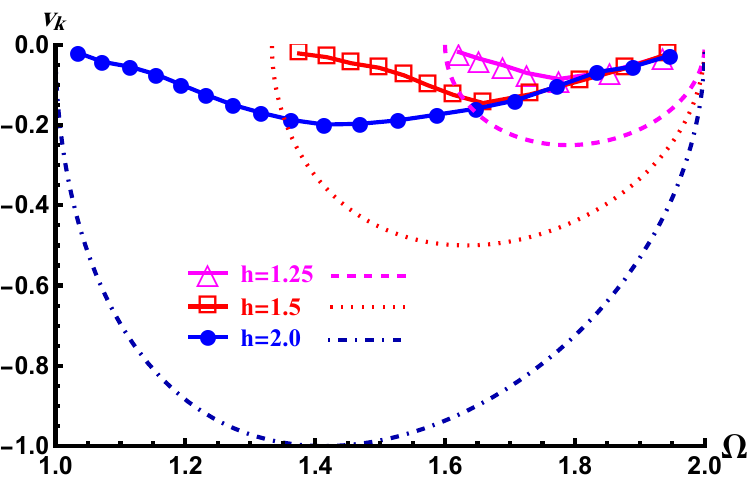}\label{fig:group_vel_shifted_vel}}
    \\
  \caption{Comparison between the phonon group velocity (dotted-dash line) and the kink velocity (markers) as functions of the driving frequency $\Omega$ for $h>1$. Panel (a) shows the comparison without Doppler correction, while panel (b) includes the Doppler-shifted group velocity. }
\label{fig:group_vel_Doppler}
\end{center}
\end{figure*}

Fig.~\ref{fig:group_vel_Doppler}(a) shows the group velocity of the phonons, given by Eq.~(\ref{eq:vgomega}), as a function of the incoming phonon frequency over the entire phonon band for strong discreteness ($h>1$), together with the kink velocities obtained in Fig.~\ref{fig:KinkVelocities1}. In this panel, the effect of the Doppler shift has not been taken into account. The group velocity shown in this panel corresponds to the positive signed wave-number branch, for which $v_g<0$ in the $h>1$ regime. This choice is useful for displaying the reversal of the dispersion slope and for comparing the location of the extrema, but the physical incident energy flux in the scattering simulations is carried by the branch directed from the source toward the kink. In Fig.~\ref{fig:group_vel_Doppler}(b), however, the Doppler correction has been included by taking the wave number from the dispersion relation Eq.~(\ref{eq:Omega}). In applying the Doppler correction, the signed wave number must be chosen consistently with the branch under consideration. As can be seen, the extrema of the Doppler-shifted group velocity nearly coincide with the frequencies at which the kink attains its maximum velocity, providing strong evidence that energy transfer is governed by the interplay between the Doppler shift and the extrema of the phonon group velocity at 
\begin{equation}
\omega_{\rm extr}=\frac{2}{\sqrt{h}}.
\end{equation}

In addition to the enhanced momentum transfer from the phonon wave packet to the kink near the extrema of the group velocity, resonance between the phonons and the kink's internal mode also plays an important role. The resonance condition can be written in terms of the Doppler-shifted phonon frequency in the kink rest frame as
\begin{equation}\label{eq:ResGenCond}
n\omega_{s}=\gamma\left[\omega(q^\star)-q^\star V_{k}\right],
\end{equation}
where $q^\star$ is the signed wave number associated with the extremum of the group velocity and $n$ is an integer indicating the resonance harmonic. For small kink velocities, $\gamma\simeq 1$, and Eq.~(\ref{eq:ResGenCond}) reduces to the usual first-order Doppler-shifted resonance condition. If one uses the positive wave-number magnitude $Q^\star=|q^\star|$, the sign of the Doppler term must be selected according to the physical branch.

For $h>1$, the extremum of the group velocity occurs at
\begin{equation}
\omega(q^\star)=\frac{2}{\sqrt{h}},
\qquad
Q^\star=\frac{2}{h}\cos^{-1}\left(\frac{1}{\sqrt{1+h}}\right).
\end{equation}
Here $Q^\star$ denotes the positive magnitude of the wave number. The corresponding signed value is either $+Q^\star$ or $-Q^\star$, depending on the branch. In the scattering geometry used here, the incident branch is the one whose group velocity points from the source toward the kink.

For the frequencies at which the kink velocity reaches its maximum, we find $\omega_s=1.03$ for $h=2.0$, $\omega_s=1.35$ for $h=1.5$ and $\omega_s=1.61$ for $h=1.25$. These values are in excellent agreement with those reported in Table~\ref{tab:shape_modes_horizontal} for the highest frequency of the shape mode. This good agreement strongly suggests that the observed maxima in the kink velocity originate from a resonant coupling between the incident phonons and the kink's internal mode, leading to an enhanced transfer of energy and momentum to the kink.

%%%%%%%%%%%%%%%%%%%%%%%%%%%%%%%%%%%%%%%%

\section{Conclusions}\label{sec:conclusions} 
We have studied the interaction of phonons with the symmetric kink initially at rest in the $\phi^4$ model in both the weakly and strongly discrete regimes, considering a system that is free of the PNP. We employed the discrete version of the model for a wide range of lattice spacings, including relatively large values of the lattice spacing, and used the energy-conservation-based numerical method introduced in \cite{Rakhmatullina2018-kz}.

Small-amplitude waves, or phonons, were generated by driving a single atom in a one-dimensional chain according to a harmonic law. The excited atom was chosen on the left side of the stationary kink, producing phonons that propagated toward the soliton. Upon interacting with the kink, the phonons were partially reflected and/or transmitted, depending on the value of the lattice spacing, $h$. For small values of $h$, the phonons were almost completely transmitted through the kink, except within a narrow frequency range near the upper edge of the phonon band. In contrast, for large values of $h$, the phonons were predominantly reflected by the kink, even though the kink itself is reflectionless.

Unlike the scattering of phonons from the asymmetric kink in the  $\phi^6$ model studied in \cite{Saadatmand:2023hqr}, where the kink can be either repelled by or attracted to the phonons, the $\phi^4$ kink is never repelled by phonons after the interaction, regardless of the value of the lattice spacing. A similar phenomenon has been observed in the scattering of wave packets from a $\phi^4$ kink  \cite{Abdelhady.IJMPA.2011,Romanczukiewicz:2005jm}, as well as in the acceleration of a breather by phonons, where an initially stationary breather acquires a constant nonzero velocity after interacting with a phonon \cite{PhysRevE.97.022217}. These observations suggest that, in the  $\phi^4$ model, the net momentum transfer from the radiation to the soliton is always directed along the propagation direction of the incident phonons, leading exclusively to negative radiation pressure. It is also shown that, in the strongly discrete regime, the kink attains its maximum velocity for frequencies near the middle of the phonon band. This behavior is a consequence of the combined effects of the extrema in the phonon group velocity and the resonance between the Doppler-shifted phonon frequency and the kink's internal mode.

For future investigations, it would be particularly interesting to consider the interaction of phonons with kinks possessing long-range tails. The scattering of large-amplitude waves from symmetric and antisymmetric kinks is also an intriguing subject for future studies. 
The dynamics then exhibit several new phenomena, including soliton acceleration, the generation of multiple soliton–antisoliton pairs, and resonant energy pumping \cite{Moradi.EPJB.2022}.

\appendix

% ==================== APPENDIX A ====================
\section{\texorpdfstring{Derivation of the Equation of Motion}{Derivation of the Equation of Motion}}
\label{app:derivation1}
	
	The action $S$ of the system in $(1+1)$ dimensions is defined as

	\begin{equation*}
		S[\phi]
		=
		\int dt \int dx \,
		\mathcal{L}(\phi,\dot{\phi},\phi^{\prime})
		=
		\int dt \int dx
		\left[
		\frac{1}{2}\dot{\phi}^{2}
		-\frac{1}{2}{\phi^{\prime}}^{2}
		-V(\phi)
		\right].
	\end{equation*}

	Applying the principle of least action, $\delta S=0$, under variations $\delta\phi(x,t)$ that vanish at the boundaries, we write

	\begin{equation*}
		\delta S
		=
		\int dt \int dx
		\left[
		\dot{\phi}\,\delta\dot{\phi}
		-
		\phi^{\prime}\,\delta\phi^{\prime}
		-
		\frac{\partial V}{\partial\phi}\,\delta\phi
		\right].
	\end{equation*}

	Using integration by parts for the temporal and spatial derivative terms yields

	\begin{equation*}
		\delta S
		=
		\int dt \int dx
		\left[
		-\ddot{\phi}
		+
		\phi^{\prime\prime}
		-
		\frac{\partial V}{\partial\phi}
		\right]
		\delta\phi.
	\end{equation*}

	With $V(\phi)=\frac{1}{2}(1-\phi^{2})^{2}$, we have $\frac{\partial V}{\partial\phi}=-2\phi(1-\phi^{2})$. Since the variation $\delta\phi(x,t)$ is arbitrary, the integrand must vanish, leading directly to

	\begin{equation*}
		\ddot{\phi}
		-
		\phi^{\prime\prime}
		-
		2\phi(1-\phi^{2})
		=
		0.
	\end{equation*}

	which is the equation of motion given in Eq.~(3).

% ==================== APPENDIX B ====================
\section{\texorpdfstring{Analytical Spectrum of the Kink Normal Modes}{Analytical Spectrum of the Kink Normal Modes}}
\label{app:derivation2}

% Reset equation numbering in this appendix: B.1, B.2, ...
\setcounter{equation}{0}
\renewcommand{\theequation}{B.\arabic{equation}}

\begingroup

In this appendix, we derive analytically the normal-mode spectrum of the
continuum $\phi^4$ kink. We consider small-amplitude perturbations around the
static kink solution
\begin{equation}
	\phi_K(x)=\tanh(x).
	\label{eq:B_kink}
\end{equation}
The perturbed field is written as
\begin{equation}
	\phi(x,t)=\phi_K(x)+\eta(x,t),
	\qquad
	|\eta(x,t)|\ll 1.
	\label{eq:B_perturbation}
\end{equation}

The equation of motion is
\begin{equation}
	\ddot{\phi}-\phi^{\prime\prime}+V'(\phi)=0,
	\label{eq:B_eom}
\end{equation}
where
\begin{equation}
	V(\phi)=\frac{1}{2}(1-\phi^2)^2.
	\label{eq:B_potential}
\end{equation}
Here, dots denote derivatives with respect to time, primes on the field
variables denote derivatives with respect to $x$, and primes on $V(\phi)$
denote derivatives with respect to the field $\phi$.

Expanding $V'(\phi_K+\eta)$ to first order in $\eta$ gives
\begin{equation}
	V'(\phi_K+\eta)
	=
	V'(\phi_K)+V''(\phi_K)\eta+\mathcal{O}(\eta^2).
	\label{eq:B_expansion}
\end{equation}
The static kink satisfies
\begin{equation}
	-\phi_K^{\prime\prime}+V'(\phi_K)=0.
	\label{eq:B_static_kink_eq}
\end{equation}
Consequently, the zeroth-order terms cancel. Retaining only the terms linear
in $\eta$, we obtain
\begin{equation}
	\ddot{\eta}-\eta^{\prime\prime}
	+V''(\phi_K)\eta=0.
	\label{eq:B_linearized}
\end{equation}

For the $\phi^4$ potential, the second derivative with respect to the field is
\begin{equation}
	V''(\phi)=6\phi^2-2.
	\label{eq:B_Vpp}
\end{equation}
Evaluating this expression on the kink background gives
\begin{align}
	V''(\phi_K)
	&=6\tanh^2(x)-2 \nonumber\\
	&=6\left[1-\operatorname{sech}^2(x)\right]-2 \nonumber\\
	&=4-6\operatorname{sech}^2(x).
	\label{eq:B_effective_potential}
\end{align}
Thus, the effective potential for the linear perturbations is
\begin{equation}
	U(x)=4-6\operatorname{sech}^2(x).
	\label{eq:B_Ux}
\end{equation}

Using the harmonic ansatz
\begin{equation}
	\eta(x,t)=\psi(x)e^{i\omega t},
	\label{eq:B_harmonic_ansatz}
\end{equation}
the linearized equation becomes
\begin{equation}
	\left[
	-\frac{d^2}{dx^2}
	+4-6\operatorname{sech}^2(x)
	\right]\psi(x)
	=
	\omega^2\psi(x).
	\label{eq:B_schrodinger}
\end{equation}
This equation is the modified P\"oschl--Teller spectral problem.

To obtain its spectrum, we rewrite Eq.~\eqref{eq:B_schrodinger} as
\begin{equation}
	\left[
	-\frac{d^2}{dx^2}
	-6\operatorname{sech}^2(x)
	\right]\psi(x)
	=
	\left(\omega^2-4\right)\psi(x).
	\label{eq:B_shifted_PT}
\end{equation}
Defining
\begin{equation}
	E=\omega^2-4,
	\label{eq:B_energy_shift}
\end{equation}
we obtain the standard P\"oschl--Teller problem
\begin{equation}
	\left[
	-\frac{d^2}{dx^2}
	-\ell(\ell+1)\operatorname{sech}^2(x)
	\right]\psi(x)
	=
	E\psi(x),
	\label{eq:B_standard_PT}
\end{equation}
where
\begin{equation}
	\ell(\ell+1)=6,
	\qquad
	\ell=2.
	\label{eq:B_ell_value}
\end{equation}

For the P\"oschl--Teller potential
$-\ell(\ell+1)\operatorname{sech}^2(x)$, the discrete eigenvalues are
\begin{equation}
	E_n=-(\ell-n)^2,
	\qquad
	n=0,1,\ldots,\ell-1.
	\label{eq:B_PT_eigenvalues}
\end{equation}
For $\ell=2$, this relation yields two normalizable bound states,
corresponding to $n=0$ and $n=1$.

\subsection*{B.1. Translational Mode:
\texorpdfstring{$\omega_0=0$}{omega0=0}}

For $n=0$, Eq.~\eqref{eq:B_PT_eigenvalues} gives
\begin{equation}
	E_0=-(2-0)^2=-4.
	\label{eq:B_E0}
\end{equation}
Since $E=\omega^2-4$, we have
\begin{equation}
	\omega_0^2-4=-4.
\end{equation}
Therefore,
\begin{equation}
	\omega_0^2=0,
	\qquad
	\omega_0=0.
	\label{eq:B_zero_frequency}
\end{equation}

The corresponding eigenfunction is proportional to the derivative of the kink:
\begin{equation}
	\psi_0(x)\propto\phi_K'(x)
	=\operatorname{sech}^2(x).
	\label{eq:B_zero_mode_unnormalized}
\end{equation}
This result also follows directly from translational invariance. Hence, $\operatorname{sech}^2(x)$ is the zero mode of the kink and the normalization integral is

\begin{equation}
	\int_{-\infty}^{\infty}
	\operatorname{sech}^4(x)\,dx
	=
	\frac{4}{3}.
	\label{eq:B_zero_norm_integral}
\end{equation}
The normalized translational mode is therefore
\begin{equation}
	\psi_0(x)
	=
	\frac{\sqrt{3}}{2}\operatorname{sech}^2(x).
	\label{eq:B_zero_mode_normalized}
\end{equation}
This zero-frequency mode is the Goldstone mode associated with the continuous
translational symmetry of the continuum kink.

\subsection*{B.2. Internal Shape Mode:
\texorpdfstring{$\omega_1=\sqrt{3}$}{omega1=sqrt3}}

For $n=1$, Eq.~\eqref{eq:B_PT_eigenvalues} gives
\begin{equation}
	E_1=-(2-1)^2=-1.
	\label{eq:B_E1}
\end{equation}
Using $E=\omega^2-4$, we obtain
\begin{equation}
	\omega_1^2-4=-1.
\end{equation}
Thus,
\begin{equation}
	\omega_1^2=3,
	\qquad
	\omega_1=\sqrt{3}.
	\label{eq:B_shape_frequency}
\end{equation}

The corresponding bound-state eigenfunction is odd under $x\to-x$:
\begin{equation}
	\psi_1(x)\propto
	\tanh(x)\operatorname{sech}(x),
	\label{eq:B_shape_mode_unnormalized}
\end{equation}
and the normalization integral is
\begin{equation}
	\int_{-\infty}^{\infty}
	\operatorname{sech}^2(x)\tanh^2(x)\,dx
	=
	\frac{2}{3}.
	\label{eq:B_shape_norm_integral}
\end{equation}
Therefore, the normalized internal mode is
\begin{equation}
	\psi_1(x)
	=
	\sqrt{\frac{3}{2}}\,
	\operatorname{sech}(x)\tanh(x).
	\label{eq:B_shape_mode_normalized}
\end{equation}
This is the localized internal, or shape, mode of the continuum $\phi^4$ kink.

\endgroup

 \section*{Acknowledgments}

The research was supported by the Higher Education and Science Committee of MESCS RA,  grant No. 25IRF/2-1C008.

\bibliographystyle{apsrev}
\bibliography{refs}

\end{document}